\PassOptionsToPackage{hyphens}{url} 
\documentclass[%
 aip,
 amsmath,amssymb,
preprint,%
]{revtex4-1}
\usepackage{dcolumn,mathptmx,setspace}
\usepackage[colorinlistoftodos]{todonotes}
\usepackage[utf8]{inputenc} 
\usepackage[T1]{fontenc}    
\DeclareMathOperator*{\loggrad}{log-grad}

\usepackage[utf8]{inputenc} 
\usepackage[noend]{algpseudocode}

\usepackage{amssymb,amsmath,mathrsfs,stmaryrd,amsthm,mathtools,amsfonts,graphicx,hyperref,url,booktabs,nicefrac,comment,microtype,tikz-cd,lingmacros,nccmath,tree-dvips, bbm, bm, etoolbox,algorithm,caption,subcaption,color,enumitem,relsize,soul}

\usepackage{amsmath}

\begin{document}

\preprint{AIP/123-QED} 
\title[Efficiency of communities and financial markets during the 2020 pandemic]{Efficiency of communities and financial markets during the 2020 pandemic}

\author{Nick James}
\affiliation{ 
School of Mathematics and Statistics, University of Melbourne, Victoria, 3010, Australia}%
\author{Max Menzies}
\email{max.menzies@alumni.harvard.edu}
\affiliation{%
Yau Mathematical Sciences Center, Tsinghua University, Beijing, 100084, China}%

\date{July 2 2021}
\begin{abstract}

This paper investigates the relationship between the spread of the COVID-19 pandemic, the state of community activity, and the financial index performance across 20 countries. First, we analyze which countries behaved similarly in 2020 with respect to one of three multivariate time series: daily COVID-19 cases, Apple mobility data and national equity index price. Next, we study the trajectories of all three of these attributes in conjunction to determine which exhibited greater similarity. Finally, we investigate whether country financial indices or mobility data responded quicker to surges in COVID-19 cases. Our results indicate that mobility data and national financial indices exhibited the most similarity in their trajectories, with financial indices responding quicker. This suggests that financial market participants may have interpreted and responded to COVID-19 data more efficiently than governments. Further, results imply that efforts to study community mobility data as a leading indicator for financial market performance during the pandemic were misguided.

\end{abstract}

\maketitle

\begin{quotation}

COVID-19 has resulted in a global pandemic with severe human, social and financial costs. To mitigate the spread of new cases, governments have implemented a multi-level approach involving social distancing, business closures and movement restrictions. Given the ubiquity of mobile devices, the effects of such restrictions on community activity and movement may be observed through Apple mobility data.\cite{applemobdata} This mobility data  provides a timely representative of the current state of community movement and activity, with some limitations. In addition, it has been hypothesized that these restrictions would have a deleterious effect on the global financial system. \cite{guardian_covidfinance} This raises the question: did financial markets or community mobility, as an approximate representative of the current level of government restrictions, exhibit an earlier crash in March 2020? This may suggest whether market participants or governments responded more quickly to the pandemic. This paper addresses this by analyzing COVID-19 cases, community mobility data and financial markets in conjunction and on a country-by-country basis. We observe substantial similarity between countries' mobility and financial trajectories as a whole, and determine that markets generally reflected the pandemic more promptly than community mobility.

\end{quotation}

\section{Introduction}
\label{sec:Introduction}

Throughout the pandemic, there has been extensive debate on the merits and necessity of measures to mitigate the spread of COVID-19. Policymakers have had to balance the ramifications of business closures, movement restrictions and lockdowns against prioritizing public health. To inform the decision-making process, there has been consistent interest in studying the impact of COVID-19 cases and deaths on various measures of community and financial activity in any candidate country. While COVID-19 counts and financial data are typically available on a daily basis, traditional economic metrics such as unemployment and GDP, which may measure whether a community is thriving or in decline, are generally only available on a monthly or quarterly basis. Both financial markets and new waves of the pandemic may change at a much greater rate than updates to these data may be available.

For a more granular and up-to-date analysis during the pandemic, many investors have used alternative data sources as an indicator of the community's activity and normalcy in various countries.\cite{bloombergmobility} We follow this trend, using daily Apple mobility data as an indicator of such activity. On a country-by-country basis, this mobility data reports various community movement indicators such as walking, driving and public transport. As varying indicators are available in different locations, we average the indicators of any candidate country to form a single mobility time series for each country. We provide a more detailed description of the data in Section \ref{sec:data}.

This paper is inspired by the substantial heterogeneity that countries across the world experienced in 2020 with respect to their COVID-19 cases and deaths,  government responses, community activity, and financial market performance. Government restrictions on movement, businesses and community activity varied substantially, both from country to country and over time. Some countries acted more aggressively in closing borders \cite{bbccloseborders_2020} and implementing contact tracing programs, \cite{Koreaguardian_2020} while other nations and states did not implement sufficient restrictions or implemented them too late. \cite{nyt2020, Scally2020} These varied responses led to substantial differences in case and death counts, \cite{James2020_chaos} community activity during lockdowns and reopening, \cite{California_closing,georgia_reopening} and financial market performance throughout 2020. \cite{Jamesfincovid}

Amid this heterogeneity, studying the relationship between cases, community activity and market performance on a country-by-country basis and over time may reveal some key relationships between community and market activity during the pandemic. First, we aim to understand which countries behaved most similarly with respect to each individual characteristic. Similarity as well as differences between countries may provide the opportunity to learn from more successful countries in the future and attribute better or worse outcomes to certain responses. Next, we wish to understand which of the three data attributes behaved most similarly as a whole. This may provide insights into which factors (epidemiological,  behavioral, and financial) are likely leading vs lagging. Finally, it is worth investigating whether governments or financial markets acted more efficiently during the pandemic, and which governments acted most (and least) efficiently. We study financial markets via country equity index prices, while government actions such as restrictions on businesses and movements are related to community mobility data. Given the substantial variance among government responses, one can highlight the most (and least) responsive governments.

There has been a litany of work examining the impact of COVID-19 on various aspects of conventional financial markets. \cite{Zhang2020finance, He2020, Zaremba2020, Mnif2020, Akhtaruzzaman2020, Akhtaruzzaman2020_2, Okorie2020, Lahmiri2020,James2021} In addition to equities, significant research has addressed shifts in cryptocurrency behavior during this period. \cite{Chu2015, Lahmiri2018, James2021_crypto, Kondor2014, Bariviera2017, AlvarezRamirez2018, Stosic2019, Stosic2019_2, Manavi2020} The study of market crises, and in particular the COVID-19 market crisis, has attracted significant interest from the nonlinear dynamics and econophysics communities. \cite{Drod2001, Drod2018, Drod2019, Drod2020, Drod2020_entropy} Building upon prior studies of COVID-19 in general and analyses of mobility data,\cite{Cot2021,Praharaj2020,Kurita2021} we believe this is the first paper to simultaneously and quantitatively address the heterogeneous evolution of  community activity and financial markets among a wide collection of countries, and examine the responsiveness of community mobility vs financial markets to COVID-19.

For this purpose, we use new and existing methods to study COVID-19, mobility and financial time series on a country-by-country basis. Within epidemiology, time series analysis has been used to study a variety of diseases such as Ebola, \cite{Funk2018, Mhlanga2019} Zika, \cite{Biswas2020, Morrison2020} and more recently, COVID-19. Methods used by nonlinear dynamics researchers to study COVID-19 are numerous, including innovations based on SIR models, \cite{SIRWeinstein2020,SIRVyasarayani2020,SIRNg2020,SIRNeves2020,SIRComunian2020,SIRCadoni2020,SIRBarlow2020,SIRBallesteros2020,Gatto2021} power-law models, \cite{Manchein2020,Blasius2020,Beare2020} distance analysis, \cite{daSilva2021,James2020_nsm} network models, \cite{Shang2020, Karaivanov2020,Ge2020,Xue2020} analyses of the dynamics of transmission and contact, \cite{Saldaa2020,Danchin2021} forecasting models, \cite{Perc2020} Bayesian methods, \cite{Manevski2020} clustering \cite{Machado2020,Jamescovideu} and many other works. \cite{Ngonghala2020,Cavataio2021,Nraigh2020,Glass2020,James2021_virulence}

This paper is structured as follows. In Section \ref{sec:data}, we describe the data sources used in this paper and perform an initial analysis on the Apple mobility data. In Section \ref{sec:Clusters}, we analyze epidemiological, community mobility and financial data independently, identifying structural similarity among the 20 countries under consideration with respect to each attribute. In Section \ref{sec:Trajectory}, we study all trajectories in conjunction to determine which data attributes display more homogeneity as a whole. In Sections \ref{sec:Clusters} and \ref{sec:Trajectory}, we first use hierarchical clustering to gain insights into the structure of the collection of time series, and then follow with a closer analysis of their dynamics, using both new and existing methods. In Section \ref{sec:Offset}, we introduce a new approach to investigate whether the impact of COVID-19 was felt by mobility or financial data first. This analyzes the extent of coincidence of COVID-19 peaks and mobility and financial troughs among different countries, and may suggest whether governments or financial markets were more efficient at responding during the pandemic. We summarize our findings and limitations regarding COVID-19 cases, community mobility and financial market performance in Section \ref{sec:Discussion}.

\section{Data}
\label{sec:data}
The countries included in this study are the following: Argentina, Australia, Brazil, Canada, Denmark, France, Germany, India, Italy, Japan, Mexico, the Netherlands, New Zealand, Norway, Russia, Singapore, Spain, Switzerland, the United Kingdom (U.K.) and the United States (U.S.). Excluding China, where mobility data is not available, this list contains most of the top 20 economies in the world by nominal GDP.\cite{worldbankgdp} For each country, we analyze three time series: COVID-19 new cases, averaged Apple mobility data across available indicators, and country financial index price. As financial data is only available on weekdays (and COVID-19 and mobility data are consistently reduced on weekends), we restrict our analysis only to trading days. Our data spans 01/13/2020 to 12/30/2020, a period of 234 trading days.

Apple mobility data provides aggregated counts of requests for directions via Apple Maps for various countries and regions. Although the data records requests, rather than distances actually travelled, many analysts have used this data throughout the pandemic as a guide into people's movements. \cite{bloombergmobility} Apple mobility data is available across 63 countries, with the notable omission of China. Each country has up to three indicators - walking, driving and transit - with available indicators varying across countries. This may provide a challenge to fairly comparing different countries' mobility activity. In this section, we perform an initial analysis of the different constituent indicators of each country's mobility data. Our aim is to show high correlation between these constituent indicators. Therefore, it is appropriate to average available indicators and yield one representative mobility time series for each country. As each constituent indicator is highly correlated, this averaged data represents each individual indicator well. One can then compare the single averaged mobility time series on a country-by-country basis.

\begin{figure}
    \centering
    \includegraphics[width=0.95\textwidth]{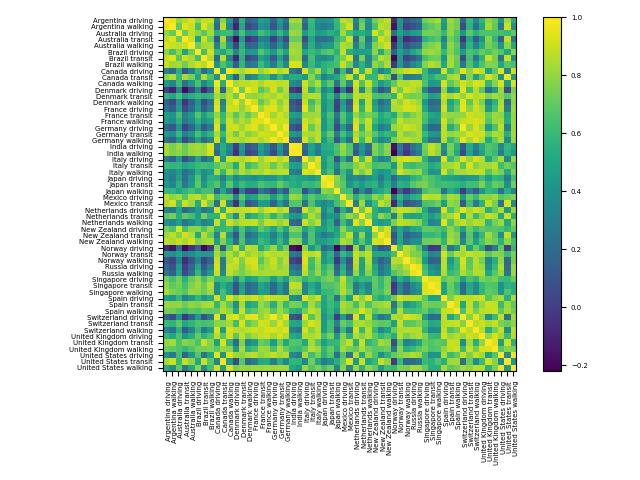}
    \caption{Correlations between time series of all available mobility indicators, walking, driving and transit. Every country has two or three available indicators. The high values around the diagonal indicate the high correlation between each country's mobility indicators, confirming Table \ref{tab:mobility_correlations}.}
    \label{fig:applecorrelations}
\end{figure}

Each country ($i=1,...,20$) has either two or three mobility indicators $\mu_i^j(t), j=1,2,3$. For each country, we compute the average of the non-trivial correlations, namely
\begin{align}
\label{eq:corrav}
    \rho^{av}_i = \begin{cases}
    \text{Corr}(\mu_i^1(t), \mu_i^2(t)) \text{ if two indicators are available;}\\
     \frac13 \left( \text{Corr}(\mu_i^1(t), \mu_i^2(t))+  \text{Corr}(\mu_i^1(t), \mu_i^3(t))+  \text{Corr}(\mu_i^2(t), \mu_i^3(t)\right) \text{ if three indicators are available.}
    \end{cases}
\end{align}
A value close to 1 indicates that all available indicators for that country are highly correlated. We record these average correlations for each country in Table \ref{tab:mobility_correlations}. In addition, we display all correlations between constituent indicators of any country in Figure \ref{fig:applecorrelations}. In Table \ref{tab:mobility_correlations}, we can see clearly that the average correlations of mobility indicators for each country are overwhelmingly high,  ranging from 0.59 to 0.99 and a median of 0.86. This is also reflected in the entries of Figure \ref{fig:applecorrelations}, where individual correlation coefficients between mobility indicators of the same country are almost all close to 1. This justifies our averaging approach, as the single averaged mobility time series for each country represents every available indicator quite well. Henceforth in this paper, we will only examine averaged mobility data in order to fairly compare country to country.

We conclude this section with a brief outline of possible bias in using the mobility data to measure community activity. If the requests that the data counts were substantially different to people's actual movements, the data may not provide the optimal depiction of community mobility. One instance of such potential bias could be people's checking travel distances to restaurants, as a means of estimating food delivery time; in such circumstances, people may not subsequently proceed with any journey. Further interpretations and limitations of this data, including existing research on the utility of mobility data, are discussed in Section \ref{sec:Discussion}.

\begin{table}
\begin{center}
\begin{tabular}{ |p{3cm}||p{3cm}|}
 \hline
 \multicolumn{2}{|c|}{Average mobility correlation} \\
 \hline
 Country & $\rho^{av}_i$ \\
 \hline
 Argentina & 0.96  \\
 Australia & 0.83  \\
 Brazil & 0.81  \\
 Canada & 0.65  \\
 Denmark & 0.83  \\
 France & 0.90  \\
 Germany & 0.93  \\
 India & 0.99  \\
 Italy & 0.88  \\
 Japan & 0.84  \\
 Mexico & 0.75  \\
 The Netherlands & 0.81  \\
 New Zealand & 0.88  \\
 Norway & 0.79  \\
 Russia & 0.96  \\
 Singapore & 0.94  \\
 Spain & 0.86  \\
 Switzerland & 0.87  \\
 United Kingdom & 0.89  \\
 United States & 0.59  \\
 \hline
 Overall mean & 0.85 \\
 Overall median & 0.86 \\
\hline
\end{tabular}
\caption{Average correlation $\rho^{av}_i$, as defined in (\ref{eq:corrav}) among individual mobility indicators for each country. Correlations are overwhelmingly high, hence we may average available indicators for each country to produce one mobility time series that is representative of each constituent indicator.}
\label{tab:mobility_correlations}
\end{center}
\end{table}

\section{Country similarity with respect to COVID-19, mobility and financial data}
\label{sec:Clusters}
In this section, we study the behaviors of COVID-19 cases, community mobility activity and financial index prices among our countries of interest. Our analysis uses data starting 01/13/2020 and finishing 12/30/2020, a period of 234 trading days. Across $n=20$ countries, let $x_i(t),\mu_i(t),p_i(t)$ be the multivariate time series of each country's new daily COVID-19 cases, averaged Apple mobility data, and daily closing equity prices, $t=0,...,T$ and $i=1,..., n$. $T=233$ gives 234 trading days.

\subsection{Methodology and overview}

First, we analyze countries' COVID-19 cases relative to each other. Considering the new case counts of a single country gives a function $\mathbf{x}_{i} \in \mathbb{R}^{T+1}$. Let $||\mathbf{x}_{i}|| = \sum^{T_1}_{t=0} |x_{i}(t)|$ be its $L^1$ norm. We can define a normalized case trajectory by $\mathbf{f}_{i} = \frac{\mathbf{x}_i}{||\mathbf{x}_i||}$. These vectors highlight the relative changes of new case counts within the entire period. We then define a \emph{trajectory distance matrix} with respect to cases, $D^C_{ij}=||\mathbf{f}_i - \mathbf{f}_j||, i,j=1,...,n$.

\begin{figure*}
    \centering
    \includegraphics[width=0.95\textwidth]{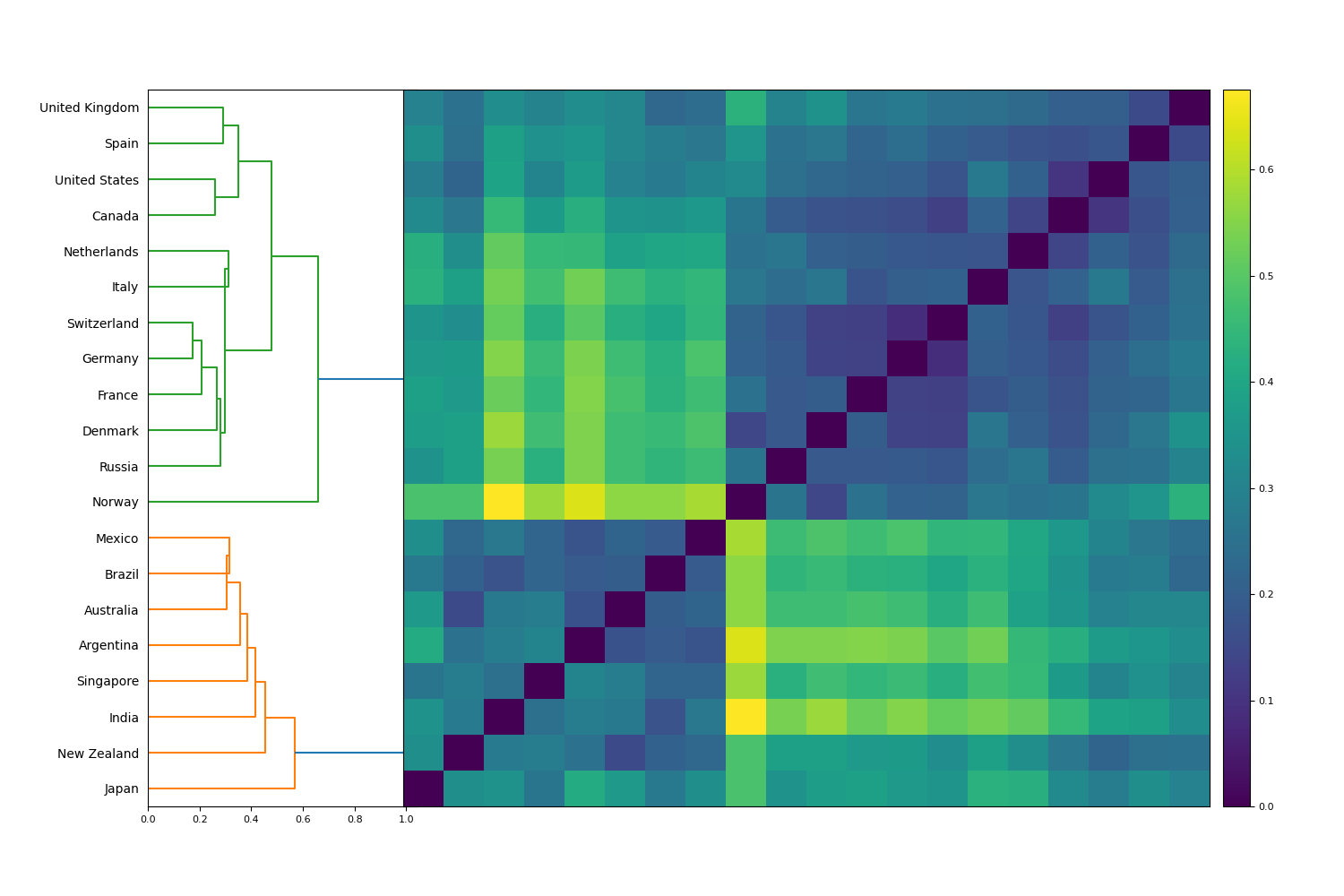}
    \caption{Mobility trajectory dendrogram, produced by hierarchical clustering on the matrix $D^M$. Countries are clustered according to their similarity in normalized trajectories of Apple mobility data.}
    \label{fig:Mobility_dendrogram}
\end{figure*}

Next, we can do the same for mobility and price data. To ensure comparably normalized trajectories, we must linearly adjust these by subtracting their minimum value. That is, let $\tilde{\mu}_i(t)=\mu_i(t) - \min_{t} \mu_i(t)$ for each country, analogously for $\tilde{p}_i(t)$. We define normalized mobility and price trajectories by  $\mathbf{g}_{i} = \frac{\mathbf{\tilde{\mu}}_i}{||\mathbf{\tilde{\mu}}_i||} $ and $\mathbf{h}_{i} = \frac{\mathbf{\tilde{p}}_i}{||\mathbf{\tilde{p}}_i||}$, respectively. Analogously as before, we define  \emph{trajectory distance matrices} with respect to mobility and prices, $D^M_{ij}=||\mathbf{g}_i - \mathbf{g}_j||$, $D^P_{ij}=||\mathbf{h}_i - \mathbf{h}_j||$. We note that COVID-19 cases exhibit their minimal value as zero, so do not need to be linearly adjusted. In Figures \ref{fig:COVID_dendrogram}, \ref{fig:Mobility_dendrogram} and \ref{fig:Financial_dendrogram}, we display hierarchical clustering on the matrices $D^C, D^M, D^P$, respectively, and proceed to interpret these results. Hierarchical clustering provides a convenient, accessible and visual means of identifying structural similarity in a collection of time series, such as which countries are most similar to each other. It provides a high-level understanding of the data, which may then be followed by more specific, targeted analysis. We supply additional analyses of the time series dynamics throughout the paper.

\subsection{Results for COVID-19 time series}

\begin{figure}
    \centering
    \includegraphics[width=0.95\textwidth]{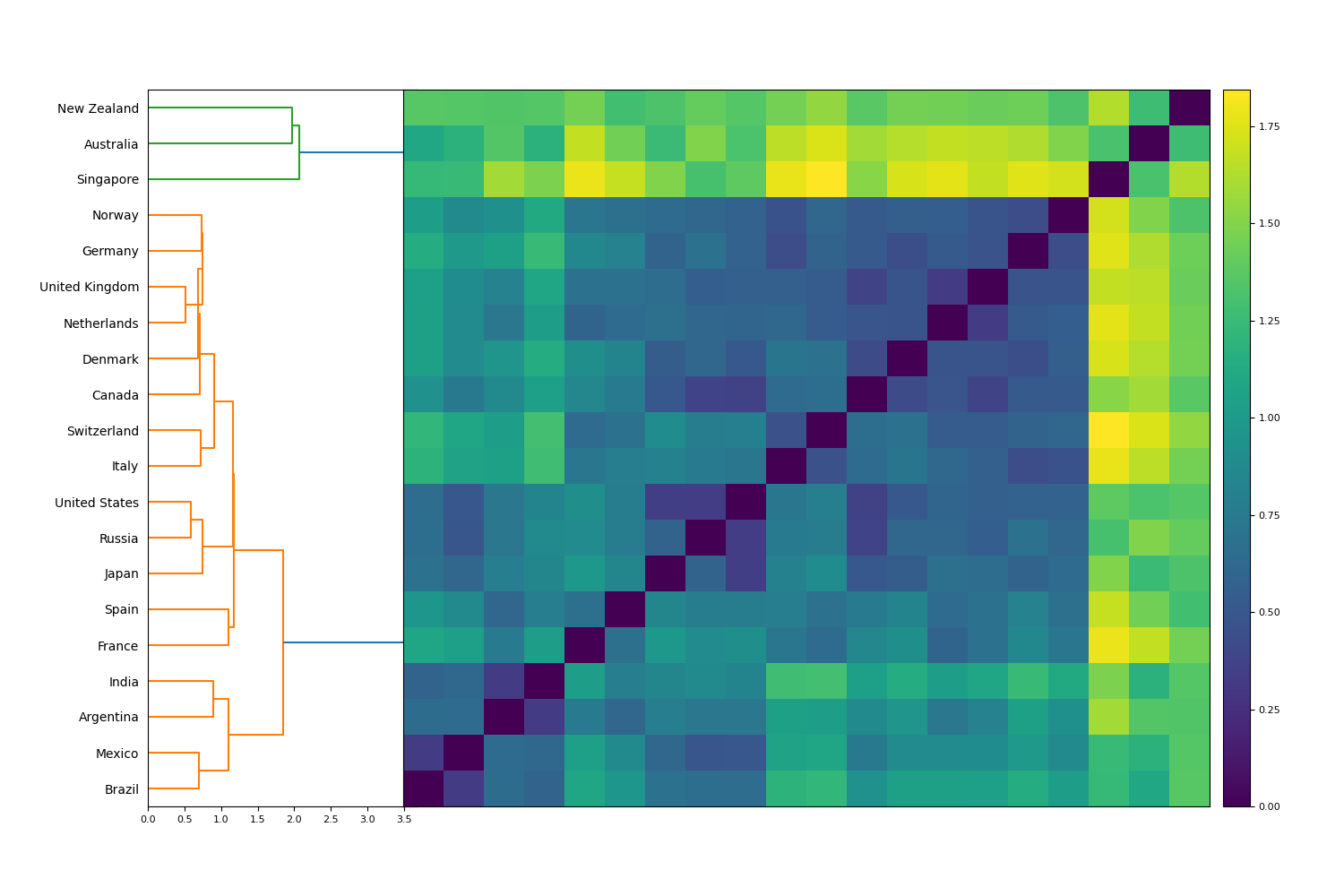}
    \caption{Cases trajectory dendrogram, produced by hierarchical clustering on the matrix $D^C$. Countries are clustered according to their similarity in normalized trajectories of new COVID-19 cases.}
    \label{fig:COVID_dendrogram}
\end{figure}

First, the cases trajectory dendrogram (Figure \ref{fig:COVID_dendrogram}) is comprised of 2 clusters, a large predominant cluster and a small cluster of outlier countries - Australia, New Zealand and Singapore. The common theme among the latter countries is their low total number of cases throughout the year and their suppressed new cases towards the end of 2020, as can be seen for Singapore in Figure \ref{fig:Singapore_cases}.  The larger cluster is characterized by new case trajectories exhibiting multiple waves of new cases, with many countries experiencing significant growth in cases towards the end of 2020, as can be seen for the U.S. in Figure \ref{fig:US_cases}.

To confirm our findings, we implement a simple analysis to quantify whether countries exhibited the majority of their cases earlier or later in the year. Given a country's new case time series $x_i(t)$, we treat the case counts as a distribution over $t=0,...,T$ and calculate its median. Specifically, we determine the minimal $T_{1/2}^{(i)}$ such that $\sum_{0 \leq t \leq T_{1/2}^{(i)}} x_i(t) \geq  \sum_{T_{1/2}^{(i)} < t \leq T} x_i(t).$ We record the values of $T_{1/2}^{(i)}$, expressed in terms of their corresponding calendar dates, in Table \ref{tab:Country_tests}. We confirm what we initially observed in Figure \ref{fig:COVID_dendrogram}, that New Zealand, Singapore and Australia have earlier values of $T_{1/2}^{(i)}$ than the remainder of the collection, explaining their nature as outliers.

\subsection{Results for mobility time series}

Next, the mobility trajectory dendrogram (Figure \ref{fig:Mobility_dendrogram}) is comprised of 2 comparably-sized clusters. The first cluster includes Argentina, Australia, Brazil, India, Japan, Mexico, New Zealand and Singapore. Mobility trajectories within this cluster experience a sharp dip early in the year, then a quick partial recovery, followed by a consistent marginal increase (or flattening off) throughout the rest of the year. These trajectories are representative of countries that experienced surges of COVID-19 cases earlier in 2020, imposed restrictions to reduce case counts, and gradually reopened their economies, such as Singapore (Figure \ref{fig:Singapore_mob}). The second cluster consists of Canada, Denmark, France, Germany, Italy, the Netherlands, Norway, Russia, Spain, Switzerland, the U.K. and the U.S.  Countries within this cluster exhibit the same sharp dip early in the year, followed by almost a complete recovery in mobility, but then declining trajectories for the rest of the year. As seen for the U.S. (Figure \ref{fig:US_mob}), this trajectory reveals a near-complete reopening and reduction of restrictions during the middle of the year, followed by further waves of cases that prompted the reinstatement of restrictions. Indeed, almost every U.S. state had reduced restrictions by May, \cite{wapo_allreopen} and subsequently had to reimpose them as cases rose once again. Interestingly, the two clusters exhibit striking geographical affinity. The first cluster consists of countries in South America and Oceania, while the second cluster comprises European and North American countries. This highlights the similarity in the timing of lockdowns among many countries in proximal geographies and could reflect common approaches among the policymakers of these respective countries.

Having observed this striking split into two clusters, we implement a new analytical method to further study this distinction quantitatively. Essentially, we wish to determine, on a country-by-country basis, the total length of time that each mobility time series was increasing vs decreasing. First, we note that every mobility time series $\mu_i(t)$ experienced its global minimum $T^\mu_i$ at approximately the same time, in March/April. The primary distinction between the two cluster behaviors occurs after this global minimum. Thus, we restrict attention to the period $t\geq T^\mu_i$.

Our approach adopts the flexible algorithmic framework of Ref. \onlinecite{james2020covidusa} to analyze the mobility time series over the period $t\geq T^\mu_i$. We begin by smoothing out each truncated (and minimum-adjusted) time series to yield a new series $\nu_i(t), t=T^\mu_i,...,T$. Then, we have two approaches, one relatively simple and one novel, to determine periods of increase and decrease. For the naive approach, let $A_i$ be the number of ``ascending'' days $t= T^\mu_i,..,T-1$ such that $\nu_i(t) \leq \nu_i(t+1)$. This is determined with the smoothed time series, so as not to be influenced by daily perturbation in the data surrounding the true signal. Our novel method, fully described in Appendix \ref{appendix:turningpoint}, identifies non-trivial turning points in each time series $\nu_i(t)$, both local maxima and minima, beginning with a minimum at $T^\mu_i$. Every day following a local minimum or maximum is counted as an increasing or decreasing period, respectively, until the next turning point is observed. With this definition, let $B_i$ be the number of days of increasing periods. Results from both the naive approach ($A_i$) and novel method ($B_i$) are recorded in Table \ref{tab:Country_tests}.

The results from our additional experiment confirm and quantify our observations from the hierarchical clustering in Figure \ref{fig:Mobility_dendrogram}, and characterize a clear difference in the dynamics of these mobility time series. 
As discussed before, hierarchical clustering identifies two distinct groups. Ordering countries in Table \ref{tab:Country_tests} by their increasing value of $B_i$ yields Spain, Russia, Denmark, the U.S., Canada, Netherlands, the U.K., Germany, Switzerland, Italy, France, and then Japan, New Zealand, Singapore, India, Mexico, Australia, Argentina, Brazil. We observe that each previously observed cluster is an unbroken interval with respect to ordering by $B_i$. This suggests that the observed cluster split in Figure \ref{fig:Mobility_dendrogram} can be largely explained by this difference in periods of increase or decrease. One cluster features countries where mobility time series exhibited decrease for much of the year, while the countries of the other cluster exhibited many more days of increase. The naive method obtains similar results, ordering the countries as Denmark, Canada, the Netherlands, Norway, Germany, the U.K., Italy, Spain, Russia, the U.S., Switzerland, Japan, France, New Zealand, Mexico, Australia, Singapore, Argentina, Brazil, and India. This almost exactly corresponds to the two clusters in Figure \ref{fig:Mobility_dendrogram}, with the switching of France and Japan. The similarity of the two approaches supports the robustness of our analysis.

\begin{table}
\begin{center}
\begin{tabular}{ |p{3cm}||p{4cm}|p{3cm}|p{3cm}|}
 \hline
 \multicolumn{4}{|c|}{Mobility and COVID-19 tests} \\
 \hline
 Country & COVID-19 median date & Mobility count $A_i$ & Mobility count $B_i$ \\
 \hline
 Argentina & 5/10/20  & 167 & 184   \\
 Australia & 27/7/20 & 139 & 179   \\
 Brazil & 27/8/20 & 167 & 185   \\
 Canada & 9/11/20 & 88 & 87  \\
 Denmark & 25/11/20  & 86 & 85  \\
 France & 28/10/20 & 123 & 119   \\
 Germany & 17/11/20& 98 & 97   \\
 India & 15/9/20 & 177 & 178   \\
 Italy & 10/11/20 & 103 & 104   \\
 Japan  & 9/11/20 & 111 & 125  \\
 Mexico & 17/9/20 & 132 & 178 \\
 The Netherlands & 2/11/20 & 94 & 89   \\
 New Zealand & 6/4/20 & 123 & 145   \\
 Norway & 5/11/20 & 95 & 83  \\
 Russia & 23/10/20 & 104 & 85   \\
 Singapore & 19/5/20 & 144 & 169   \\
 Spain & 20/10/20 & 104 & 82  \\
 Switzerland & 6/11/20  & 111 & 98 \\
 United Kingdom & 5/11/20 & 99 & 96  \\
 United States & 3/11/20 & 109 & 87 \\
\hline
\end{tabular}
\caption{COVID-19 median date and counts of increasing days in mobility time series by country, as measured by the naive ($A_i$) and novel ($B_i$) methods. These methods are described in Section \ref{sec:Clusters} with more detail in Appendix \ref{appendix:turningpoint}.}
\label{tab:Country_tests}
\end{center}
\end{table}

\subsection{Results for financial index time series}

\begin{figure*}
    \centering
    \includegraphics[width=0.95\textwidth]{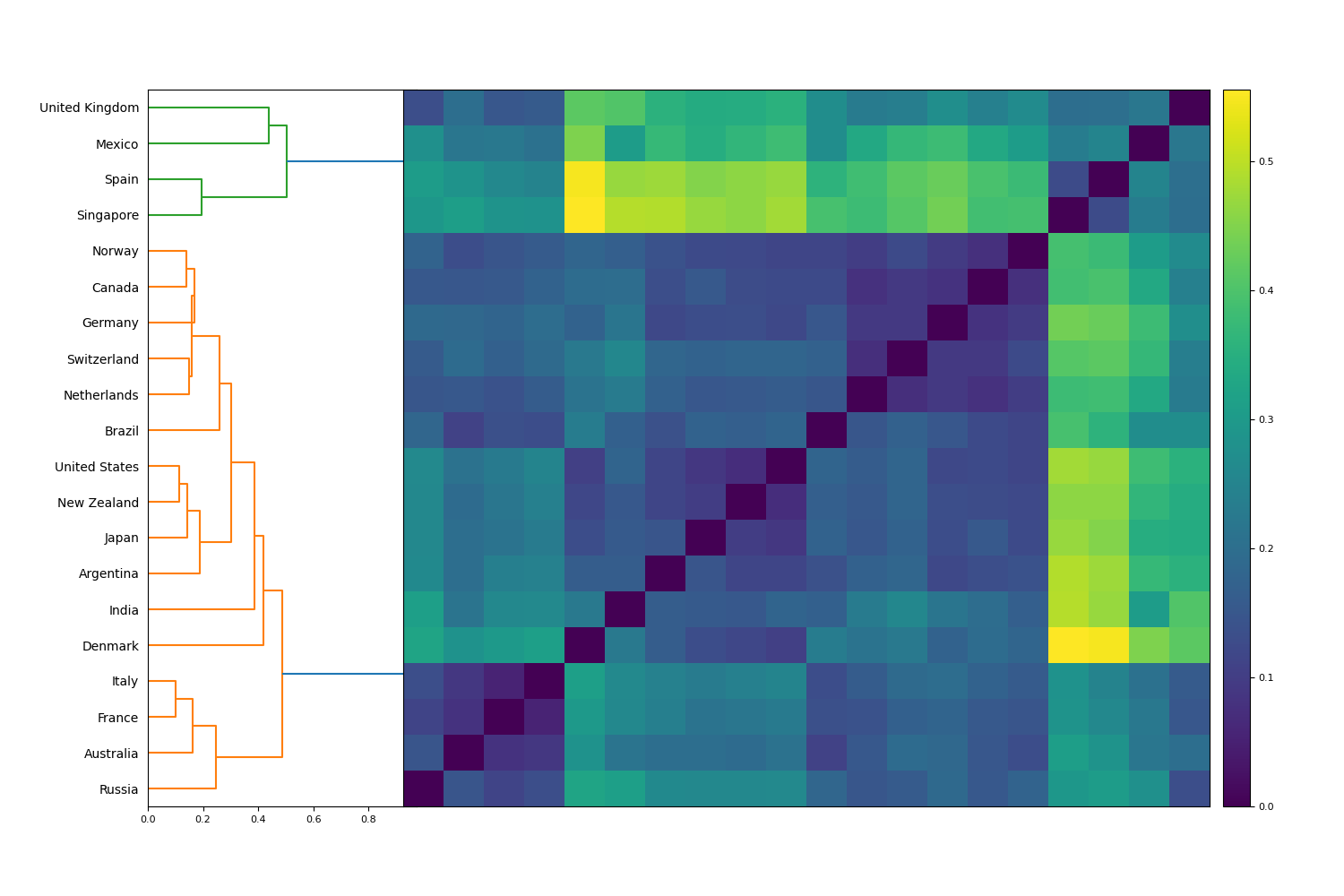}
    \caption{Prices trajectory dendrogram, produced by hierarchical clustering on the matrix $D^P$. Countries are clustered according to their similarity in normalized trajectories of financial index prices.}
    \label{fig:Financial_dendrogram}
\end{figure*}

\begin{figure*}
    \centering
    \begin{subfigure}[b]{0.48\textwidth}
        \includegraphics[width=\textwidth]{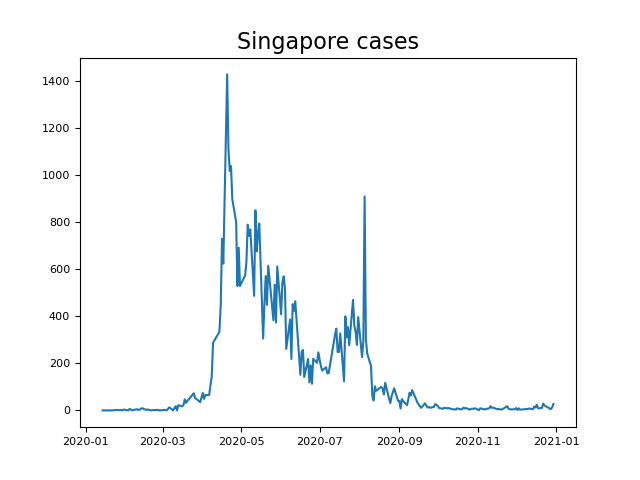}
        \caption{}
        \label{fig:Singapore_cases}
    \end{subfigure}
    \begin{subfigure}[b]{0.48\textwidth}
        \includegraphics[width=\textwidth]{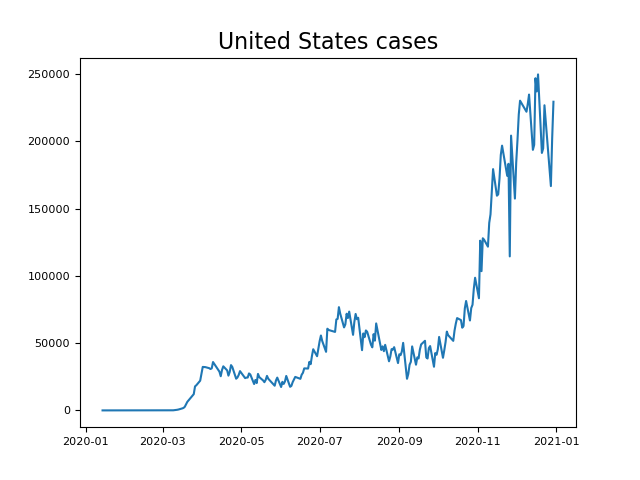}
        \caption{}
        \label{fig:US_cases}
    \end{subfigure}
    \begin{subfigure}[b]{0.48\textwidth}
        \includegraphics[width=\textwidth]{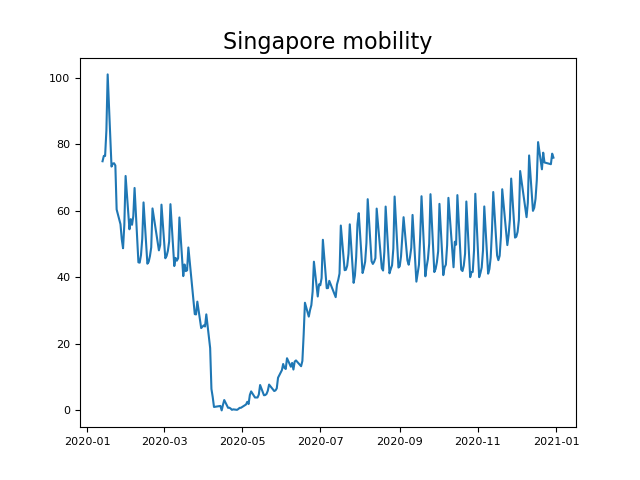}
        \caption{}
        \label{fig:Singapore_mob}
    \end{subfigure}
    \begin{subfigure}[b]{0.48\textwidth}
        \includegraphics[width=\textwidth]{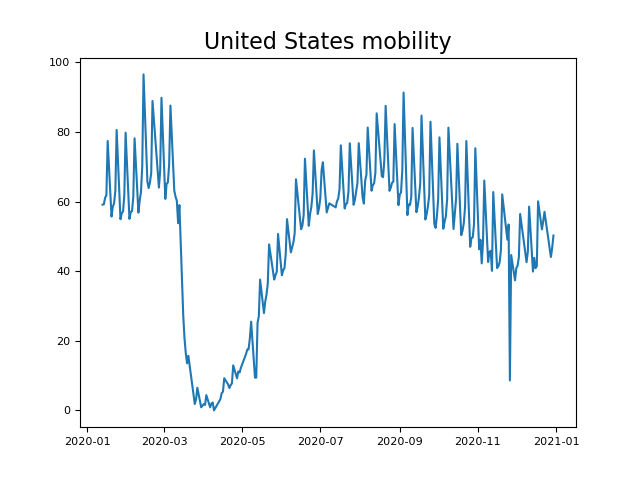}
        \caption{}
        \label{fig:US_mob}
    \end{subfigure}
    \begin{subfigure}[b]{0.48\textwidth}
        \includegraphics[width=\textwidth]{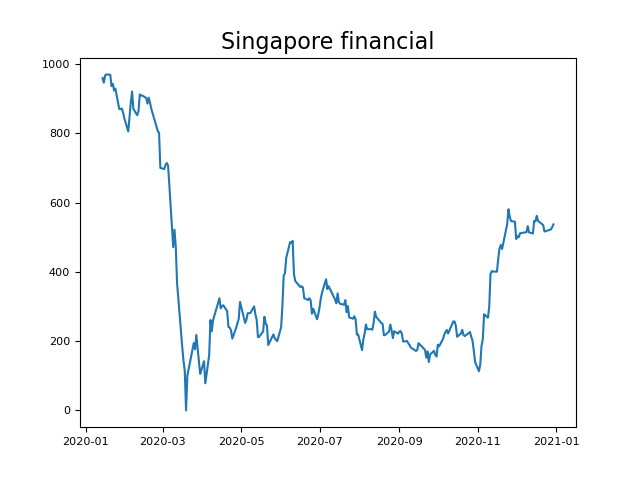}
        \caption{}
        \label{fig:Singapore_index}
    \end{subfigure}
    \begin{subfigure}[b]{0.48\textwidth}
        \includegraphics[width=\textwidth]{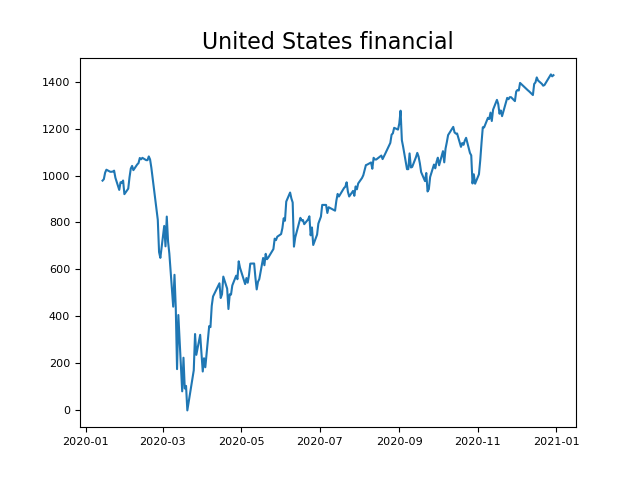}
        \caption{}
        \label{fig:US_index}
    \end{subfigure}
    \caption{Time series of COVID-19 cases $x_i(t)$ for (a) Singapore and (b) the U.S., minimum-adjusted mobility data $\tilde{\mu}_i(t)$ for (c) Singapore and (d) the U.S. and financial index $\tilde{p}_i(t)$ for (e) Singapore and (f) the U.S.}
    \label{fig:TimeSeries}
\end{figure*}

Third, the financial trajectory dendrogram (Figure \ref{fig:Financial_dendrogram}) consists of 2 clusters, a primary cluster and a smaller anomalous cluster, consisting of Mexico, Singapore, Spain and the U.K. These countries share similar financial index trajectories: there is a large drop in March, a subsequent drop until November and a moderate recovery during the latter half of November and December, as can be seen for Singapore (Figure \ref{fig:Singapore_index}). All of these countries' financial indices finish the year lower than their pre-COVID position. The dominant cluster, which consists of the remaining countries, exhibits far less uniform trajectories. Broadly speaking, countries in the dominant cluster experienced more substantial recoveries than the anomalous cluster. For instance, New Zealand and the U.S., which possess highly similar financial trajectories, both exhibit consistent gains after their precipitous drops in March, as can be seen for the U.S. in Figure \ref{fig:US_index}.

The preceding analysis highlights several noteworthy points regarding the consistency in country cluster behaviors: 
\begin{enumerate}
    \item There is little consistency in cluster structure among COVID-19 cases, mobility data and financial index performance. The size of the clusters is highly variable. However, in all dendrograms, there are 2 main clusters.
    \item There is little consistency in cluster composition. In both the COVID-19 cases and financial index dendrograms, there is a small, anomalous cluster, but limited intersection between the countries in each anomalous cluster. This highlights a lack of consistency in how countries group under different metrics.
    \item There is clearly no relationship between countries' financial index performance and their mobility or COVID-19 trajectories. The most obvious example of this is Singapore. Hierarchical clustering highlights that Singapore managed COVID-19 cases anomalously well, while its financial index performance was anomalously poor. By contrast, the U.S. leads the world in COVID-19 cases and deaths, but exhibited robust financial index performance after the initial drop in March.
\end{enumerate}

\section{Collective trajectory analysis}
\label{sec:Trajectory}
In this section, we seek to determine collective similarity among the three attributes under consideration. Specifically, we wish to measure the similarity of relative moment between new cases, changes in mobility, and market price. For this purpose, we define a symmetric $60 \times 60$ matrix $M$ that appropriately compares this similarity among the three time series. First, we define
\begin{align}
||\mathbf{x}_i||_2&=\left(\sum_{t=0}^{T} x_i(t)^2\right)^\frac{1}{2},\\
||\mathbf{\tilde{\mu}}_i||_2&=\left(\sum_{t=0}^{T} \tilde{\mu}_i(t)^2\right)^\frac{1}{2},\\ 
<\mathbf{x}_i,\mathbf{\tilde{\mu}}_j>_n&=\frac{1}{||\mathbf{x}_i||||\mathbf{\tilde{\mu}}_j||} \sum_{t=0}^{T} x_i(t)\tilde{\mu}_j(t).
\end{align}
and analogously for pairings $<\mathbf{x}_i,\mathbf{\tilde{p}}_k >,$ and $<\mathbf{\tilde{\mu}}_j,\mathbf{\tilde{p}}_k >, i,j,k=1,...,20$. The pairing $<.,.>_n$ is a normalized inner product that takes values in $[0,1]$. Two time series $f$ and $g$ have maximal similarity $<f,g>_n=1$ if and only if $f=kg$ for a constant $k>0$ - this is an important property for our application. The matrix $M$ uses this pairing to measure similarity between all three attributes and is defined as follows:
\begin{align}
M_{ij}=\begin{cases}
<x_i(t),x_j(t)>_n \text{ if } 1\leq i,j \leq 20,\\
<\tilde{\mu}_{i-20}(t),\tilde{\mu}_{j-20}(t)>_n \text{ if } 21\leq i,j \leq 40,\\
<\tilde{p}_{i-40}(t),\tilde{p}_{j-40}(t)>_n \text{ if } 41\leq i,j \leq 60,\\
<x_i(t),\tilde{\mu}_{j-20}(t)>_n \text{ if } 1\leq i \leq 20, 21 \leq j \leq 40,\\
<x_i(t),\tilde{p}_{j-40}(t)>_n \text{ if } 1\leq i \leq 20, 41 \leq j \leq 60,\\
<\tilde{\mu}_{i-20}(t),\tilde{p}_{j-40}(t)>_n \text{ if } 21\leq i \leq 40, 41 \leq j \leq 60.
\end{cases}
\end{align}
First, we explain why we use this pairing rather than the more standard correlation or distance correlation metrics. Two new case time series $\mathbf{x}_i$ and $\mathbf{x}_j$ have maximal correlation or distance correlation 1 if $\mathbf{x}_i=k\mathbf{x}_j+b$ for constants $k>0$ and $b$. This property is unsuitable for uncovering similarity between time series regarding movement relative to itself. For example, if the cases of one country $\mathbf{x}_i$ change linearly from 0 to 100 and return to 0, this would be maximally correlated with the cases of another country $\mathbf{x}_j$ that change linearly from 1000 to 1100 and return to 1000 over the same time period. Clearly, these countries differ substantially in their changes in new cases relative to themselves, with $\mathbf{x}_j$ exhibiting a much flatter trajectory and less movement relative to itself. The same reasoning applies for mobility and price trajectories.

\begin{figure*}
    \centering
        \includegraphics[width=\textwidth]{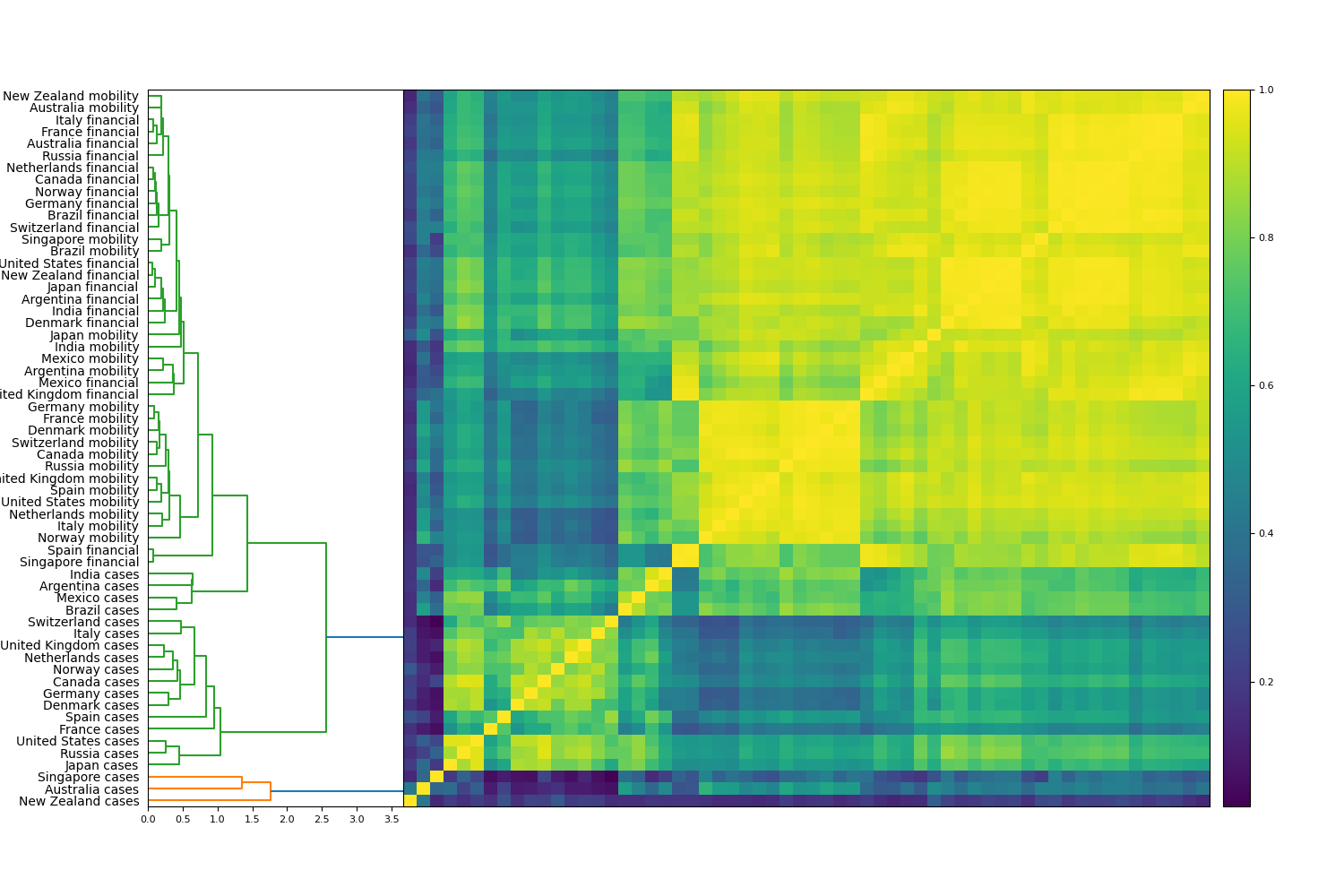}
        \caption{Hierarchical clustering on the matrix $M$, defined in Section \ref{sec:Trajectory}, measures the similarity of relative moment between cases, mobility data and financial indices of 20 countries.}
    \label{fig:InnerProduct_dendrogram}
\end{figure*}

We perform hierarchical clustering on $M$ in Figure \ref{fig:InnerProduct_dendrogram}, and observe substantial structure in this dendrogram. Financial and mobility trajectories are highly similar both across different countries and to each other, while COVID-19 case trajectories are significantly less similar. Both the mobility and financial index trajectories exhibit similar shapes: there is generally a sharp drop around March, followed by a subsequent recovery in community mobility and market price later in the year. COVID-19 cases, on the other hand, exhibit more heterogeneous trajectories. Indeed, due to the infectious nature of the virus and differing government approaches, COVID-19 case counts may change substantially over time and form different waves of the outbreak, constituting multiple peaks and troughs throughout the year.

We can quantify the level of heterogeneity between the three data attributes by further examining the matrix $M$. Restricting $M$ to its first $20 \times 20$ rows and columns yields a matrix $M_{\text{case}}$ that describes the similarity between countries' case trajectories. We can similarly produce $M_{\text{mob}}$ and $M_{\text{fin}}$. Given an $n \times n$ matrix $A$, define its (normalized) $L^1$ norm by $||A||=\frac{1}{n^2}\sum_{i,j=1}^{n} |A_{ij}|$. Applied to $M_{\text{case}}$, $M_{\text{mob}}$ and $M_{\text{fin}}$, these norms quantify the total (average) similarity within these matrices. We see that $||M_{\text{mob}}||=0.93$, indicating an average similarity of 0.93 between all the mobility trajectories, while $||M_{\text{fin}}||=0.77$ and $||M_{\text{case}}||=0.61.$ This reveals the strongest self-similarity between mobility trajectories, followed by financial and then case trajectories. Indeed, mobility trajectories exhibit substantial homogeneity, with only slight differences described between the 2 clusters of Figure \ref{fig:Mobility_dendrogram}.

Figure \ref{fig:InnerProduct_dendrogram} also contains numerous insights pertaining to specific countries. For example, the case trajectories of Australia, New Zealand and Singapore are identified as highly anomalous, as observed previously in Figure \ref{fig:COVID_dendrogram}. Singapore and Spain's financial trajectories, while somewhat similar to other financial indices, are revealed as moderately anomalous, as discussed in Figure \ref{fig:Financial_dendrogram}. Australia and New Zealand have highly similar mobility trajectories, reflecting the two government's similarly aggressive responses in limiting public movement. \cite{AusNZsimilar} Even with the substantial similarity between mobility and financial trajectories as a whole, there is no tendency for trajectories to be grouped based on country. For example, neither Singapore nor the U.S.' COVID-19 cases, mobility or financial index trajectories cluster similarly.

\section{Temporal analysis of community and financial response to COVID-19}
\label{sec:Offset}

\begin{figure*}
    \centering
    \includegraphics[width=0.95\textwidth]{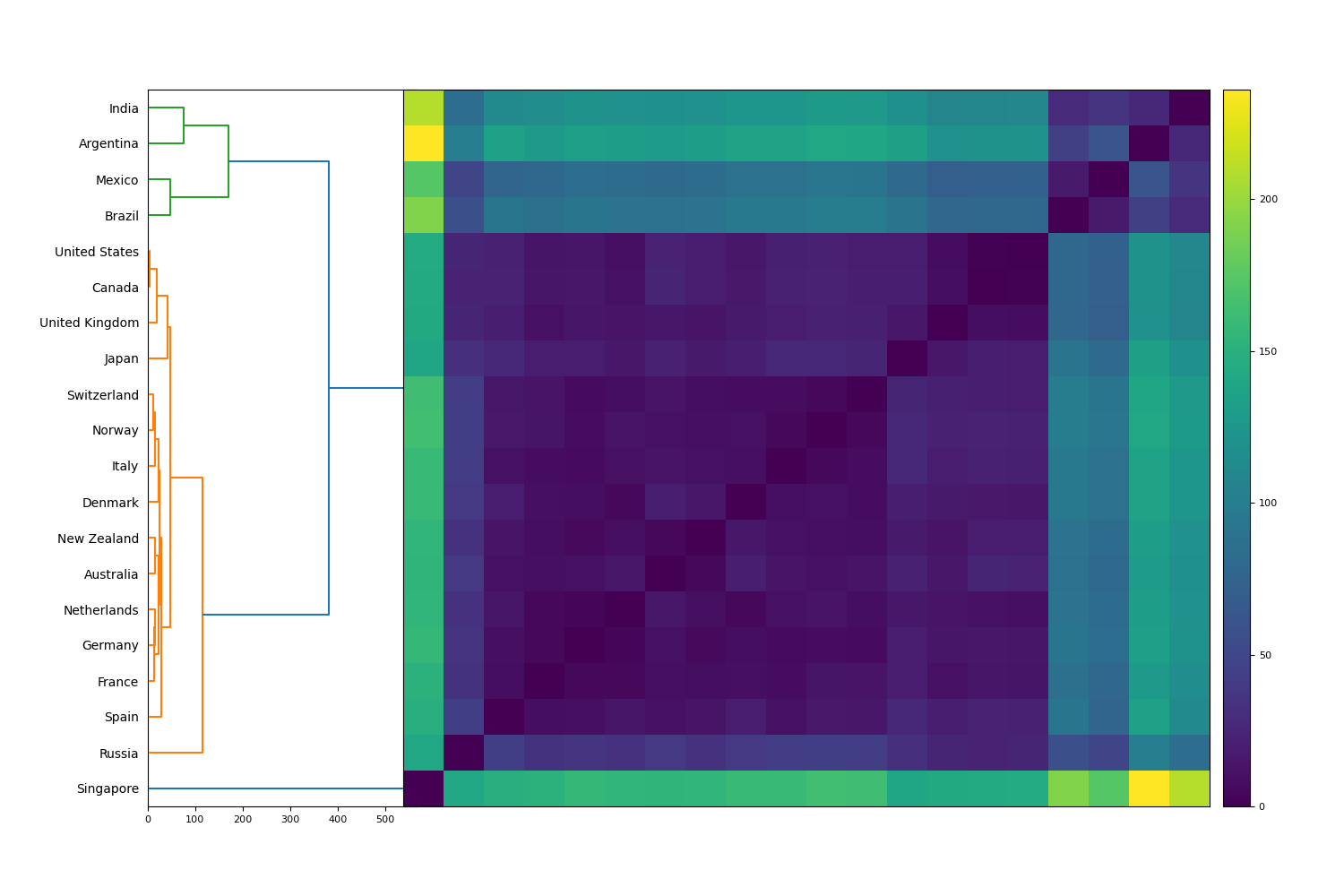}
    \caption{Hierarchical clustering on the matrix $D^{cp}$, defined in Section \ref{sec:Offset}, measures the similarity between the first critical points of cases, mobility and financial index time series. Singapore is the only country whose global (smoothed) financial minimum $T^P_i$ did not coincide with its first trough, which we have chosen to distinguish here.}
    \label{fig:3_way_trajectory_dendrogram}
\end{figure*}

\begin{figure*}
    \centering
    \begin{subfigure}[b]{0.49\textwidth}
        \includegraphics[width=\textwidth]{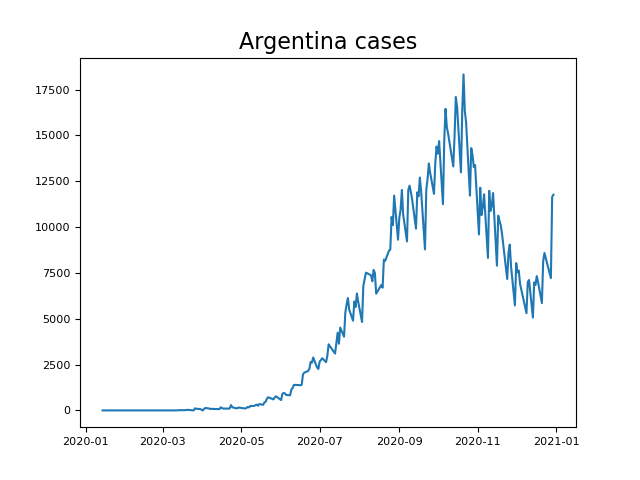}
        \caption{}
        \label{fig:Argentina cases}
    \end{subfigure}
    \begin{subfigure}[b]{0.49\textwidth}
        \includegraphics[width=\textwidth]{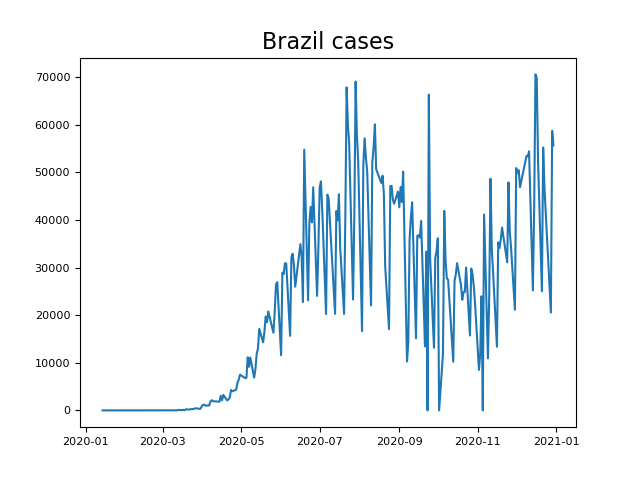}
        \caption{}
        \label{fig:Brazilcases}
    \end{subfigure}
\begin{subfigure}[b]{0.49\textwidth}
        \includegraphics[width=\textwidth]{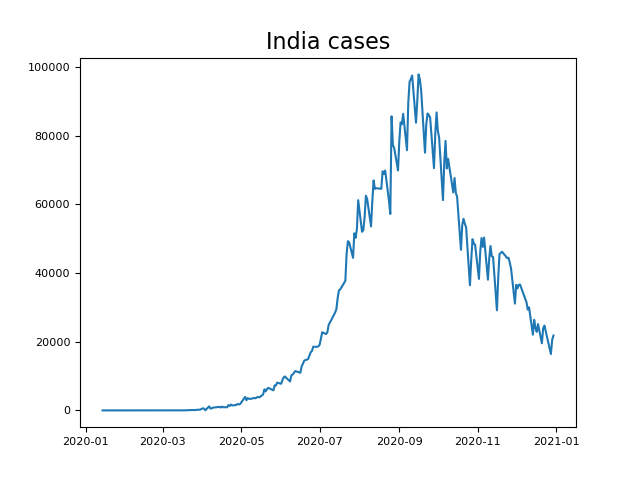}
        \caption{}
        \label{fig:Indiacases}
    \end{subfigure}
\begin{subfigure}[b]{0.49\textwidth}
        \includegraphics[width=\textwidth]{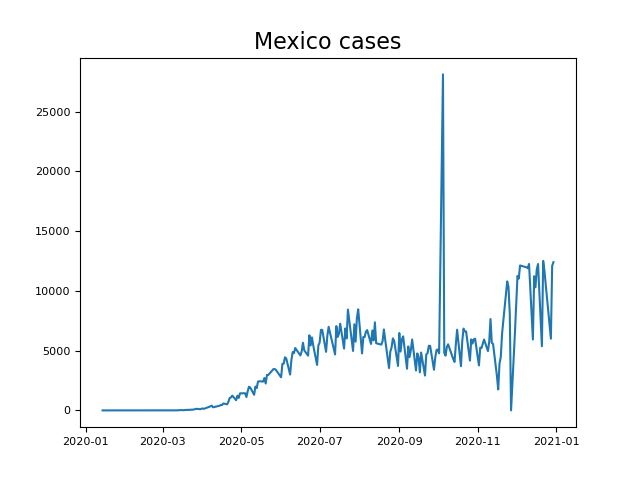}
        \caption{}
        \label{fig:Mexicocases}
    \end{subfigure}
    \caption{COVID-19 new case time series for (a) Argentina (b) Brazil (c) India (d) Mexico. Brazil and Mexico exhibit their first peak in smoothed cases around the start of August, while India and Argentina exhibit their first peak even later, in September and October, respectively.}
   \label{fig:Argentinaetc}
\end{figure*}

In this section, we analyze the collective similarity between mobility data and financial indices in more detail. In particular, we investigate mobility activity or financial indices responded first to the pandemic - and whether this varied among countries. Our analytical approach uses the same three multivariate time series of the previous section as proxy data sets: COVID-19 cases, averaged mobility data and country financial indices. Observing these time series, we can see that most countries experienced their first peak in COVID-19 cases, signifying the height of the first wave, at a similar time as the first troughs in mobility and price data. We pay particular attention to these first critical points in the time series. In order to remove outlier points, we smooth out all time series in the proceeding analysis. Applying the flexible algorithmic framework of Ref. \onlinecite{james2020covidusa}, detailed in Appendix \ref{appendix:turningpoint}, let $T^C_i$ be the time at which the first peak (local maximum) in (smoothed) cases occurred for country $i=1,...,n$. Analogously, let $T^M_i$ and $T^P_i$ be the times at which the global minimum in smoothed mobility and prices occurred, for $i=1,...,n$. For every mobility and price time series other than Singapore's price index, this global minimum coincides with the first trough (local minimum) in smoothed mobility and prices, respectively. That is, except for Singapore as an anomaly, the times $T^M_i$ and $T^P_i$ are the first troughs in (smoothed) mobility and prices, and are thus appropriate to compare with $T^C_i$. We can then define a \emph{critical point distance matrix} between countries as follows:
\begin{align}
 D^{cp}_{ij}=|T^C_i - T^C_j|+|T^M_i - T^M_j| + |T^P_i - T^P_j|, i,j=1,...,n.
\end{align}
We perform hierarchical clustering on this matrix $ D^{cp}$ in Figure \ref{fig:3_way_trajectory_dendrogram}.
We may also define an \emph{internal critical distance} between these critical points for any candidate country. With $q>0$ a fixed parameter, we define
\begin{align}
d^{\text{int}}_i=\left(\frac{|T^C_i - T^M_i|^q+|T^M_i - T^P_i|^q + |T^P_i - T^C_i|^q}{3}\right)_.^{\frac{1}{q}}
\end{align}
These values are appropriately normalized for comparison as the parameter $q$ varies. If $q=1$, this simplifies to $\frac{2}{3}( \max\{T^C_i,T^M_i,T^P_i\} - \min\{T^C_i,T^M_i,T^P_i\})$. When $q<1$, this internal distance becomes smaller when the middle critical value is closer to one of the end points and thus rewards (via a smaller total value) less equally distributed points. When $q>1$ the reverse occurs - $d_i^{\text{int}}$ penalizes equally distributed points. The values of $d_i^{\text{int}}$ for all 20 countries and varying values of $q$ are recorded in Table \ref{tab:3_way_table}.

\begin{table}
\begin{center}
\begin{tabular}{ |p{3cm}||p{2cm}|p{2cm}|p{2cm}|}
 \hline
 \multicolumn{4}{|c|}{Internal critical distance $d^{\text{int}}$ by country for three parameters $q$} \\
 \hline
 Country & $q=1/2$ & $q=1$ & $q=2$ \\
 \hline
 Argentina & 66.7 & 82  & 97.7  \\
 Australia & 9.1 & 9.3  & 9.9 \\
 Brazil & 44.5 & 52.7 & 61.8 \\
 Canada & 9.0 & 9.3  &  9.9 \\
 Denmark & 2.7 & 4 &  4.9 \\
 France & 2.2 &  3.3 &  4.1 \\
 Germany & 2.7 & 4 & 4.9 \\
 India & 52.3 & 66 & 79.3 \\
 Italy & 1.3 & 2 & 2.4 \\
 Japan & 12.6 & 13.3 & 14.5  \\
 Mexico & 34.5 & 42  & 49.9 \\
 The Netherlands & 4.5  & 4.7 & 5.0\\
 New Zealand & 7.2  & 8  & 9.1\\
 Norway & 3.8 & 4 & 4.3 \\
 Russia & 19.3 & 20 & 21.3 \\
 Singapore & 45.8 & 68.7 & 84.1 \\
 Spain & 5.2 & 6 & 7.0 \\
 Switzerland & 4.1 & 4.7 & 5.4  \\
 United Kingdom & 5.2   & 6  & 7.0 \\
 United States & 8.4 & 8.7 & 9.3\\
\hline
\end{tabular}
\caption{Internal critical distances between critical points $T^C_i, T^M_i, T^P_i$ for every country and various parameters $q$.}
\label{tab:3_way_table}
\end{center}
\end{table}

Viewed in conjunction, Figure \ref{fig:3_way_trajectory_dendrogram} and Table \ref{tab:3_way_table} reveal notable anomalies in the timing of the critical points of COVID-19 cases, mobility data and financial indices. The countries with the five greatest internal distances, as recorded in Table \ref{tab:3_way_table}, for any value of $q$, are Argentina, Brazil, India, Mexico and Singapore. All of these also appear as outliers in Figure \ref{fig:3_way_trajectory_dendrogram}. Singapore is a lone outlier due to a financial minimum occurring late in the year, much later than any other country. If we were to change the definition of $T^P_i$ to be the first trough, Singapore would disappear as an outlier. We choose to document it as it is, noting its unique status in having a second and more substantial minimum in (smoothed) index price later in the year (Figure \ref{fig:Singapore_index}). Argentina, Brazil, India and Mexico occur as an outlier cluster in Figure \ref{fig:3_way_trajectory_dendrogram}. All four of these emerging economies exhibit similar early troughs in mobility activity and financial indices when compared with the other countries under consideration. They are anomalous due to their much later first peaks in COVID-19 cases, as seen in Figure \ref{fig:Argentinaetc}. Brazil and Mexico, which form a subcluster pair, exhibit their first peak in smoothed cases near the start of August (Figures \ref{fig:Brazilcases} and \ref{fig:Mexicocases} respectively). India and Argentina, which also form a subcluster pair, exhibit their first peaks in cases in September and October, respectively (Figures \ref{fig:Indiacases} and \ref{fig:Argentina cases}, respectively). That is, COVID-19 cases were consistently and uniformly rising throughout almost the whole year in these countries. 

On the other hand, the countries with the lowest internal critical distances (both for $q=1/2$ and $q=1$) are, in order, Italy, France, Denmark, Germany, Norway, Switzerland, the Netherlands, Spain, and the U.K. Once again, we observe substantial similarity in these countries - all are wealthy European nations. Together with similarly developed countries Australia and New Zealand, these form a subcluster of close similarity in Figure \ref{fig:3_way_trajectory_dendrogram}. This reveals that all these countries behaved similarly relative to COVID-19 counts, mobility data and financial indices early in the year, with an early COVID-19 peak, mobility trough and financial trough in March. Of course, the latter half of 2020 featured substantial differences between, for example, Australia and New Zealand and the aforementioned European countries, as shown in Figure \ref{fig:COVID_dendrogram} and analyzed further in Table \ref{tab:Country_tests}.

Nowhere in the preceding analysis do we determine whether $T^M_i$ or $T^P_i$ occurred first; we only consider the absolute distances in these values between each other and other countries. We conclude this section with an inner product-based optimization framework that computes an offset between the mobility and financial time series for each country. This considers all data points in the time series (not just the critical points) to identify whether mobility data lags financial performance or vice versa. Specifically, we wish to determine the offset $\tau \in \{-30,-29,...,-1,0,1,...,29,30\}$ for each country $i$ that most appropriately aligns the function $\tilde{p}_i(t)$ with the translated function $\tilde{\mu}_i(t+\tau)$. If $\tau>0$, this means the values of $\tilde{\mu}_i(t+\tau)$ most resemble those of $\tilde{p}_i(t)$, and hence that mobility data is lagging prices. We allow $\tau$ to be both positive or negative, so our approach remains flexible to the possibility of either time series lagging the other. We define our normalized inner product in full generality as follows: let $\tilde{p}$ and $\tilde{\mu}$ be the minimum-adjusted price and mobility time series, respectively, of a given country (see Section \ref{sec:Clusters}) and define
\begin{equation}
\label{eq:bigthing}
\mathlarger{\frac{ \mathlarger{\sum}_{0\leq s,t \leq T, t-s = \tau} \tilde{p}(s)\tilde{\mu}(t)}{ \left(\mathlarger{\sum}_{\max(0,-\tau)\leq s \leq \min(T-\tau,T)} \tilde{p}(s)^2\right)^\frac{1}{2} \left(\mathlarger{\sum}_{\max(0,\tau)\leq t \leq \min(T,T+\tau)} \tilde{\mu}(t)^2\right)_.^\frac{1}{2}  }}
\end{equation}
When $\tau>0$, this can be more easily rewritten as 
\begin{align}
<\tilde{p}(0:T-\tau),\tilde{\mu}(\tau:T)>_n =
\frac{\tilde{p}(0)\tilde{\mu}(\tau)+...+\tilde{p}(T-\tau)\tilde{\mu}(T)}{(\tilde{p}(0)^2+...+\tilde{p}(T-\tau)^2)^\frac{1}{2}(\tilde{\mu}(\tau)^2+...+\tilde{\mu}(T)^2)^\frac{1}{2}},
\end{align}
using the notation of Section \ref{sec:Trajectory}. When $\tau<0$, let $\sigma=-\tau$, then (\ref{eq:bigthing}) can be written as
\begin{align}
<\tilde{p}(\sigma:T),\tilde{\mu}(0:T-\sigma)>_n =
\frac{\tilde{p}(\sigma)\tilde{\mu}(0)+...+\tilde{p}(T)\tilde{\mu}(T-\sigma)}{(\tilde{p}(\sigma)^2+...+\tilde{p}(T)^2)^\frac{1}{2}(\tilde{\mu}(0)^2+...+\tilde{\mu}(T-\sigma)^2)^\frac{1}{2}},
\end{align}
using the same notation. We document the optimal offset values in Table \ref{tab:OffsetsTable}. We record the absolute value $|\tau|$ of each offset, as well as the interpretation of whether mobility lags (when $\tau>0$) or price lags (when $\tau<0$). 

\begin{table}
\begin{center}
\begin{tabular}{ |p{3cm}||p{3cm}|p{3.3cm}|}
 \hline
 \multicolumn{3}{|c|}{Offsets between mobility and price trajectories} \\
 \hline
 Country & Lag & Absolute offset $|\tau|$ \\
 \hline
 Argentina & Mobility lags & 6 \\
 Australia & Mobility lags & 7 \\
 Brazil & Mobility lags & 7 \\
 Canada & Mobility lags & 10 \\
 Denmark & Price lags & 3 \\
 France & Mobility lags & 3 \\
 Germany & Mobility lags & 3 \\
 India & Mobility lags & 9 \\
 Italy & Price lags & 3 \\
 Japan & Mobility lags & 18 \\
 Mexico & Mobility lags & 15 \\
 The Netherlands & Mobility lags & 3 \\
 New Zealand & Mobility lags & 11 \\
 Norway & Mobility lags & 4 \\
 Russia & Mobility lags & 11 \\
 Singapore & Mobility lags & 7 \\
 Spain & Mobility lags & 2 \\
 Switzerland & Price lags & 2 \\
 United Kingdom & Mobility lags & 10 \\
 United States & Mobility lags & 7 \\
\hline
\end{tabular}
\caption{Determined offsets between mobility and financial data for each of 20 countries, together with the identification of which attribute is lagging.}
\label{tab:OffsetsTable}
\end{center}
\end{table}

We summarize several findings revealed therein:
\begin{enumerate}
    \item In general, mobility trajectories lag financial index trajectories. In fact, only 3 of the 20 countries possess trajectories where their mobility activity reflects COVID-19 case data more quickly than the respective financial index. 
    
    \item There is significant variability in the offset among countries. The range of offsets is 21 days, with Japan's mobility data lagging 18 days behind its price index, while Denmark and Italy's mobility data lead 3 days ahead of their price indices. Spain and Switzerland have the smallest absolute offsets of only 2 days, indicating substantial concurrence of mobility and financial data.
    
\item In general, the extent of mobility lagging suggests that communities were overwhelmingly slower to respond to the impact of COVID-19 than financial markets. In particular, using community activity and mobility data to predict financial movement is likely to be systematically flawed, as financial markets reacted well before such data was available. We believe that the mobility data reflects government restrictions as well, because imposed restrictions, such as movement restrictions or business closures, may have an immediate impact on people's queries regarding walking, driving and transit to nearby destinations in their community. We further discuss such interpretations in Section \ref{sec:Discussion}.
\end{enumerate}

\section{Conclusion}
\label{sec:Discussion}

In this paper, we have analyzed 20 countries with significant financial economies, examining their Apple mobility and financial data in relation to their COVID-19 counts. The mobility data is a useful representative for the current level of community activity and hence, we will argue, the current level of government restrictions imposed on the community. In our analysis, we have demonstrated that there is little consistency in cluster membership when examining the trajectories of COVID-19 cases, mobility activity and financial index performance. Explored in more detail in Figure \ref{fig:3_way_trajectory_dendrogram}, we saw there is no tendency for specific countries' features to be clustered together. This is most clearly evidenced by Singapore, who managed its COVID-19 case numbers particularly well but had uniquely weak performance in its financial index. By contrast, the United States experienced huge COVID-19 case and death counts, while the national equity index experienced strong performance after the market crash in March. 

In addition, we investigated the varying heterogeneity of COVID-19 cases, mobility and financial data across our collection of countries. Both qualitatively, in the descriptions of Figures \ref{fig:COVID_dendrogram}, \ref{fig:Mobility_dendrogram} and \ref{fig:Financial_dendrogram} as well as quantitatively, by examining $||M_{\text{case}}||, ||M_{\text{mob}}||$ and $||M_{\text{fin}}||$, we have observed the highest homogeneity in mobility trajectories. Even where mobility trajectories split into two clusters in Figure \ref{fig:Mobility_dendrogram}, the difference in behavior between the two clusters was relatively mild, and could largely be explained by differing periods of increase and decrease in Table \ref{tab:Country_tests}. On the other hand, COVID-19 cases exhibited the most substantial heterogeneity between countries, consistent with the natural history of COVID-19 varying substantially around the world.

Examining the mobility and price trajectories in a temporal analysis revealed that most developed countries experienced a small lag between the first trough in mobility and financial data, around the same time as their first peak in cases. However, emerging economies Argentina, Brazil, India and Mexico experienced their first peak in cases much later, especially for Argentina and India. The finding that mobility data lagged financial data is significant for two reasons. First, it shows that an analysis of such mobility data to predict financial market performance during the pandemic (both in the past and going forward) is likely to be misguided, as financial time series consistently led the mobility time series. Secondly, it suggests that governments, whose restrictions were arguably the main cause of dips in mobility data, were less efficient in reacting and imposing restrictions than market participants.


There are numerous strengths, possible interpretations, and limitations to this article, and our results could be of interest to various different entities, including economists, behavioral researchers, and governments. One potential interpretation of the mobility data is a proxy for certain aspects and sectors of the community's \emph{real economy}, as distinct from a country's financial economy. Indeed, researchers have consistently aimed to study the impact of the pandemic on the real economy, but have experienced difficulties, as traditional economic metrics such as GDP and unemployment are typically available on an infrequent basis. During the pandemic, various market participants have attempted to use mobility data as a means to detect a revival in the real economy.\cite{bloombergmobility} This makes sense, as community activity, while not associated with all sectors or aspects of the economy, may be closely associated with certain sectors such as retail, hospitality, and transport.

Next, Apple mobility data has a relationship with the current level of government restrictions imposed on a community. Indeed, Apple mobility data has been used previously by researchers to measure the current impact of and adherence to social distancing, one of the most commonly enacted government regulations.\cite{Cot2021} We do not study government response as an explicit variable, as this is difficult to quantify for the purposes of time series analysis. Instead, we rely on mobility data as a related, but not perfect, proxy. Previous research\cite{Praharaj2020} has shown a close relationship between stay-at-home orders and mobility, demonstrating that the decline in mobility data observed in March/April actually preceded government restrictions in certain locations. That is, government response was even slower than mobility data (which we have shown, in turn, was slower than financial markets). The same work also found that trip requests in certain locations increased after lockdown interventions were lifted. Thus, there is certainly a considerable association between mobility data and government restrictions, though they are not identically aligned.

Therefore, our analysis of the dynamics of financial indices and mobility data may complement existing research well, showing broadly speaking that financial markets preceded the change in community mobility, which preceded government actions, at least in imposing restrictions in March/April of 2020. Our research may therefore be of use to market analysts, those interested in community activity, and governments. Other strengths of this article include the fact that mobility is analyzed across 19 countries, rather than existing studies that have studied just several cities, and mobility data is analyzed in comparison with financial data. In addition, we have determined not just broad trends between mobility data and financial indices, but determined which time series lagged and by how much for every country under consideration.

Drawing such interpretations from Apple mobility data and comparing its dynamics with financial indices carries numerous limitations and avenues of improvement in future research. First, we lack a control group, so can make no claims on the causality of certain time series moving before others.\cite{Pearl2009} Thus, we can only suggest, but not definitively state, that COVID-19 cases impacted financial markets before community activity. Second, financial markets are a complex system, and may contain dynamics that our analysis may not be privy to. Third, in order to compare different countries, we have had to average mobility indicators. While we have demonstrated these were highly correlated, this may reduce specific insights. For example, walking and driving are likely to be more associated with the retail sector, while the transit indicator is much more representative of the transport sector of the economy. The strength of the hospitality sector may be associated with both the walking and driving indicators (representing the patronage of locals) and the transit indicator (representing the patronage of tourists). Fourth, Apple mobility data includes any lookup of directions, and does not correspond exactly to community movement. These all make the Apple mobility data an imperfect representative of community mobility or the current level of government restrictions.

Financial markets are notorious for their complexity, with many variables and interrelationships making it near impossible to extract clear causal relationships. This is often the reason many statistical learning algorithms, when applied to problems in financial markets, exhibit low signal:noise ratios. It is virtually impossible to engineer parametric models with complexity sufficient to model the true data generating process of financial market behaviors, without over-fitting. Our approach allows for reasonable inference on which time series led and lagged in a complex, high dimensional system, where devising sufficiently complex algorithms is a difficult task.  Our approach yields key descriptive findings, even though a full description of the dynamics and interplay of mobility and financial time series would be extremely challenging and is an ongoing goal of the field.

Future research could address these limitations in various ways. Rather than averaging all available mobility indicators, a closer analysis of individual indicators could provide more insights. For example, public transportation is more directly responsive to shifts in government restrictions, while car transportation may closer represent autonomous community movement. Our analysis of country financial indices could be replaced with a finer analysis on a sector-by-sector basis to determine the sectors with the most responsive market participants. In particular, some financial sectors may be more responsive to or impactful on mobility data, including individual indicators. Alternatively, one could study the behavior of other asset classes and the underlying relationships with mobility data. Asset classes such as fixed income, cryptocurrencies, and commodities such as gold and oil may all exhibit varied relationships with mobility data. Finally, instead of analyzing daily mobility data, one could examine other daily data that represents the current activity of the real economy or level of government restrictions, or even incorporate traditional economic metrics that are less frequently available. Due to the infrequent nature of traditional metrics, market participants are always on the lookout for such indicators.

Overall, this manuscripts presents several findings regarding the structure and dynamics of COVID-19 cases, Apple mobility data and financial indices across 19 countries. We have combined existing techniques such as hierarchical clustering with new analytical methods to provide visual inference regarding countries' similarity relative to these attributes, followed by more quantifiable details. Our principal findings are that mobility time series exhibited more homogeneity among countries than case or financial time series, and usually lagged behind financial indices. Our paper complements existing research and offers several avenues for further study.

\section*{Data availability}
COVID-19 case counts are sourced from Our World in Data, \cite{worldindata2020} 
Apple mobility data are sourced from Apple, \cite{applemobdata}
 and financial data are sourced from Bloomberg.

\appendix
 
\section{Turning point methodology}
\label{appendix:turningpoint}

In this section, we provide more details for the identification of turning points of a smoothed time series, which we utilize twice in the body of the paper. In Section \ref{sec:Clusters}, we apply this procedure to smoothed truncated mobility time series to calculate the parameter $B_i$; in Section \ref{sec:Offset}, we apply it to a new case time series $x_i(t)$ to isolate the first non-trivial trough $T^C_i$. In both cases, smoothing is necessary due to noise and irregularities in the data sets. We apply a Savitzy-Golay filter, which combines polynomial interpolation and a moving average computation, to perform the smoothing.

The initial steps are as follows. Given a mobility time series $\mu_i(t)$, we identify its global minimum $T^\mu_i$, which occurs in March/April 2020 for every country. We form the minimum-adjusted time series $\tilde{\mu}_i(t)$ and truncate it to the period $t=T^\mu_i,...,T$. This series is non-negative everywhere, beginning with a zero value at its initial value $t=T^\mu_i$. Then, we perform the smoothing to produce a smoothed truncated time series $\nu_i(t) \in \mathbb{R}_{\geq 0}$, beginning with an initial value of zero. Given a new case time series $x_i(t)$, we simply apply smoothing immediately to yield a smoothed time series $\hat{x}_i(t)$. We proceed by describing our methodology on both time series: let $\hat{z}(t)$ be an arbitrary smoothed time series, beginning with a minimal value of 0 at its initial value, which we label $t=0$.

Following Ref. \onlinecite{james2020covidusa}, we apply a two-step algorithm to $\hat{z}(t)$. The first step produces an alternating sequence of troughs and peaks, beginning with a trough at $t=0$, where the time series exhibits its global minimum. The second step refines this sequence according to chosen conditions and parameters. The most important conditions to initially identify a peak or trough, respectively, are the following:
\begin{align}
\label{baddefnpeak}
\hat{z}(t_0)&=\max\{\hat{z}(t): \max(0,t_0 - l) \leq t \leq \min(t_0 + l,T)\},\\
\label{baddefntrough}\hat{z}(t_0)&=\min\{\hat{x}(t): \max(0,t_0 - l) \leq t \leq \min(t_0 + l,T)\},
\end{align}
where $l$ is a parameter to be chosen. Following Ref. \onlinecite{james2020covidusa}, we select $l=17$, originally inspired by the 14-day incubation period of the virus.\cite{incubation2020} Defining peaks and troughs according to this definition alone has some flaws, such as the potential for two consecutive peaks.

Instead, we implement an inductive procedure to choose an alternating sequence of peaks and troughs. Suppose $t_0$ is the last determined peak. We search in the period $t>t_0$ for the first of two cases: if we find a time $t_1>t_0$ that satisfies (\ref{baddefntrough}) as well as a non-triviality condition $\hat{z}(t_1)<\hat{z}(t_0)$, we add $t_1$ to the set of troughs and proceed from there. If we find a time $t_1>t_0$ that satisfies (\ref{baddefnpeak}) and  $\hat{z}(t_0)\geq \hat{z}(t_1)$, we ignore this lower peak as redundant; if we find a time $t_1>t_0$ that satisfies (\ref{baddefnpeak}) and  $\hat{z}(t_1) > \hat{z}(t_0)$, we remove the peak $t_0$,  replace it with $t_1$ and continue from $t_1$. A similar process applies from a trough at $t_0$. 

At this point, the time series is assigned an alternating sequence of troughs and peaks. However, some turning points are immaterial and should be removed. Let $t_1<t_3$ be two peaks, necessarily separated by a trough. We select a parameter $\delta=0.2$, and if the \emph{peak ratio}, defined as $\frac{\hat{z}(t_3)}{\hat{z}(t_1)}<\delta$, we remove the peak $t_3$. If two consecutive troughs $t_2,t_4$ remain, we remove $t_2$ if $\hat{z}(t_2)>\hat{z}(t_4)$, otherwise remove $t_4$. That is, if the second peak has size less than $\delta$ of the first peak, we remove it.

Finally, we use the same \emph{log-gradient} function between times $t_1<t_2$, defined as
\begin{align}
\label{loggrad}
   \loggrad(t_1,t_2)=\frac{\log \hat{z}(t_2) - \log \hat{z}(t_1)}{t_2-t_1}.
\end{align}
The numerator equals  $\log(\frac{\hat{x}(t_2)}{\hat{x}(t_1)})$, a "logarithmic rate of change." Unlike the standard rate of change given by $\frac{\hat{z}(t_2)}{\hat{z}(t_1)} -1$, the logarithmic change is symmetrically between $(-\infty,\infty)$. Let $t_1,t_2$ be adjacent turning points (one a trough, one a peak). We choose a parameter $\epsilon=0.01$;  if
\begin{align}
    |\loggrad(t_1,t_2)|<\epsilon,
\end{align}
that is, the average logarithmic change is less than 1\%, we remove $t_2$ from our sets of peaks and troughs. If $t_2$ is not the final turning point, we also remove $t_1$.

We conclude with an alternating sequence of peaks and troughs, beginning with a trough at the initial value. From here, we may extract the parameters that we use in the body of the text. First, consider the minimum-adjusted smoothed truncated time series $\nu_i(t)$. Let 
\begin{align}
    A_i&=|\{t: T^\mu_i\leq t \leq T-1, \nu_i(t)\leq \nu_i(t+1) \}|;\\
    B_i &= |\{t: T^\mu_i\leq t \leq T-1, \text{ the turning point immediately preceding $t$ is a trough} \}|.
\end{align}
We refer to $A_i$ and $B_i$ as the naive and novel approaches to count the number of increasing days. The second approach counts periods of more essential increase, whereas $A_i$ is more vulnerable to insubstantial day-to-day fluctuations in the data, even after smoothing. Both these counts are recorded in Table \ref{tab:Country_tests}. Finally, consider a (smoothed) new case time series $\hat{x}_i(t)$. We simply define $T^C_i$ to be the first trough after $t=0$, which exists for every country. This marks the end of the first wave.

\bibliography{__references}
\end{document}